\documentclass[prl,aps,twocolumn,superscriptaddress,floatfix,citeautoscript,showpacs]{revtex4-1}
\usepackage{graphicx,rotating,subfigure,amsmath,amsfonts,amssymb,delarray,color}
\usepackage{hyperref}
\usepackage{xcolor}
\usepackage{soul}
\usepackage{dsfont}
\usepackage[T1]{fontenc}
\usepackage{physics}
\usepackage{multirow}
\usepackage{comment}

\newcommand{\e}{\text{e}}

\def\12{\frac{1}{2}}

\begin{document}
\bibliographystyle{apsrev}

\title{Evidence for unbounded growth of the number entropy\\ in many-body localized phases}

\author{Maximilian Kiefer-Emmanouilidis}
\affiliation{Department of Physics and Research Center OPTIMAS, University of Kaiserslautern, 67663 Kaiserslautern, Germany}
\affiliation{Department of Physics and Astronomy, University of Manitoba, Winnipeg R3T 2N2, Canada}
\author{Razmik Unanyan}
\affiliation{Department of Physics and Research Center OPTIMAS, University of Kaiserslautern, 67663 Kaiserslautern, Germany}
\author{Michael Fleischhauer}
\affiliation{Department of Physics and Research Center OPTIMAS, University of Kaiserslautern, 67663 Kaiserslautern, Germany}
\author{Jesko Sirker}
\affiliation{Department of Physics and Astronomy, University of Manitoba, Winnipeg R3T 2N2, Canada}

\date{\today}

\begin{abstract}
We investigate the number entropy $S_N$---which characterizes
particle-number fluctuations between subsystems---following a quench
in one-dimensional interacting many-body systems with potential
disorder. We find evidence that in the regime which is expected to show
many-body localization (MBL) and where the entanglement entropy grows as $S\sim
\ln t$ as function of time $t$, the number entropy grows as
$S_N\sim\ln\ln t$, indicating continuing particle transport at a very
slow rate. We demonstrate that this growth is consistent with a
relation between entanglement and number entropy recently established
for non-interacting systems.
\end{abstract}

\maketitle

\paragraph{Introduction.---$\!\!\!\!$}
The time dependence of the entanglement entropy $S(t)$ after a quantum
quench offers insights into the dynamics of quasi-particles and the
influence of conservation laws. Well studied are quenches starting
from a product state in clean lattice models with short-range hoppings
and interactions. In this case, the generic picture is one of
quasi-particles propagating through the system with a velocity bounded
by the Lieb-Robinson velocity $v_\textrm{LR}$
\cite{LiebRobinson, Bravyi2006,Eisert2006}. The entanglement entropy is then proportional to the entangled region
created by the quasi-particle excitations. For a subsystem with volume
$\ell^d$ in $d$ dimensions, this leads to $S
\sim\ell^{d-1}t$ for times $v_\textrm{LR}t\ll \ell$ and a volume-law saturation, 
$S
\sim \ell^d$, at times $v_\textrm{LR}t\gg\ell$. This picture
has been confirmed in free scalar field theories
\cite{CotlerHertzberg} and in one-dimensional systems which are 
conformally invariant \cite{CalabreseCardy}. An obvious exception from
a linear increase of the entanglement entropy after a quench and from
a volume-law scaling at long times are disordered non-interacting
systems in an Anderson localized (AL) phase \cite{Anderson}. In this case, the
spreading of excitations is limited to the localization length
$\xi_{\textrm{loc}}$ leading to an area law,
$S
\sim\ell^{d-1}\xi_{\textrm{loc}}$, instead of a
volume law at long times. The increase of the entanglement entropy
after the quench is therefore bounded \cite{ZhaoSirker2019}.

In recent years, the question of localization in the presence of
interactions---termed {\it many-body localization} (MBL)---has
attracted renewed interest
\cite{AleinerAltshuler,ZnidaricProsen,PalHuse,NandkishoreHuse,AltmanVoskReview,
AbaninRev2019}. For the spin-$1/2$ Heisenberg chain with local
magnetic fields drawn from a box distribution, numerical data appear
consistent with a transition from an ergodic phase at small disorder
to a non-ergodic MBL phase at strong disorder
\cite{PalHuse,Luitz1,Luitz2}. One of the hallmarks of MBL as compared to AL
is the unbounded logarithmic growth of $S$ after a quench
\cite{ZnidaricProsen,BardarsonPollmann,AndraschkoEnssSirker}. Recently,
evidence for $S\sim\ln t$ has also been obtained in an experiment on
cold atomic gases \cite{LukinRispoli}. Here a quench in a
one-dimensional Aubry-Andr\'e model of interacting bosons was studied
with single atom resolution. In such systems where the total particle
number (or similarly the total magnetization) is conserved, the von
Neumann entropy can be split into two parts, $S= S_N + S_c$
\cite{WisemanVaccaro,Rakovszky2019,LukinRispoli,Bonsignori2019}. Here
\begin{equation}
\label{Snum}
S_N = -\sum_n p(n)\ln p(n)
\end{equation}
is the number entropy with $p(n)$ the probability of finding $n$ atoms
in the considered subsystem (also referred to as charge
\cite{Rakovszky2019} or fluctuation entropy \cite{Bonsignori2019}). The
configurational entropy $S_c$ then contains the contributions to
entanglement due to configurational correlations. This splitting of
$S$ is not only useful from an experimental perspective because $p(n)$
can be determined by single-site resolution atomic imaging
\cite{LukinRispoli} but also offers further insights 
into questions of localization and ergodicity. Very recently, we have
shown that in any non-interacting fermionic system $S^{(2)}\propto
\exp(S^{(2)}_N)$ where $S^{(2)}$ is the second R\'enyi entropy and $S^{(2)}_N$ 
the corresponding number entropy. I.e., a growth in the entanglement
entropy is always accompanied by a logarithmically slower growth in
the number entropy  \cite{KieferUnanyan1}.

An exception to this picture of correlated dynamics of entanglement and number
entropies is expected to occur in many-body localized (MBL) phases. Here $S_N$ is believed to saturate after a
quantum quench, indicating localization, while $S$ continues to grow
in time.  It has been argued that MBL systems are described at long
times by effective Hamiltonians \cite{HuseNandkishore,SerbynPapic}
\begin{equation}
\label{Heff}
H = \sum_i \varepsilon_i \eta_i + \sum_{i,j} J_{ij} \eta_i \eta_j + \cdots
\end{equation}
with exponentially many local conserved charges $[H,\eta_i]=0$, random
energies $\varepsilon_i$, and amplitudes $J_{ij}$ which decay
exponentially with distance between these charges. As a consequence of
the coupling terms $\sim J_{ij}$, a region of length $\ell$ will
become entangled over time $t\sim\e^\ell$. Since the entanglement
entropy is extensive, one then expects $S\sim \ell\sim \ln t$
\cite{Serbyn2013} consistent with the numerical and experimental
observations. If Eq.~\eqref{Heff} is a valid effective discription of
MBL phases of matter, then the increase in entanglement at long times
is entirely due to the continuing buildup of configurational
entanglement $S_c$. Since the conserved charges $\eta_j$ are local,
the number entropy $S_N$ has to be bounded, reflecting the expected
localized and non-ergodic character of this phase. On the other hand,
the experimental data for the number entropy in
Ref.~\onlinecite{LukinRispoli} appear to show a slow increase,
although a detailed analysis of the number entropy as a function of
system size and disorder strength has not yet been
performed. Furthermore, it has recently been suggested that
paradigmatic models expected to show MBL phases might ultimately be
ergodic at very long times
\cite{SuntajsBonca,Znidaric2018}.

These recent results motivate us to investigate the number entropy in
systems believed to show MBL. In this letter we provide evidence that the
picture of MBL phases based on effective Hamiltonians \eqref{Heff}
might be incomplete. For all system sizes and times we can access
numerically, we find that the number entropy grows as $S_N\sim\ln\ln
t$ even at strong disorder and does not show any signs of
saturating. We, furthermore, present evidence that the relation
$S^{(2)}\propto\exp(S^{(2)}_N)$, proven for free fermionic systems in
\cite{KieferUnanyan1}, also appears to hold in the interacting case,
both in the ergodic and in the MBL phase, with proportionality factors
renormalized by interactions and disorder.

\paragraph{Number entropies.---$\!\!\!\!$}
If we split a one-dimensional system $S$ into two parts, $A$ and $B$,
then the R\'enyi entanglement entropies are given by
\begin{equation}
\label{Sent}
S^{(\alpha)} = (1-\alpha)^{-1}\ln\tr \rho^\alpha_A
\end{equation} 
where $\rho_{A}$ is the reduced density matrix of the considered
subsystem. The von-Neumann entanglement entropy is given by
$S
\equiv S^{(1)}=\lim_{\alpha\to 1} S^{(\alpha)}$. If
the total particle number is conserved, then we can write
$S^{(\alpha)} = (1-\alpha)^{-1}\ln(\sum_n
p^\alpha(n)\tr\rho^\alpha_A(n))$ where $\rho_A(n)$ is the block of the
reduced density matrix with particle number $n$ normalized such that
$\tr\rho_A(n)=1$. If there is only a single configuration for each $n$
then $\tr\rho^\alpha_A(n)=\tr\rho_A(n)=1$. We thus call
$S^{(\alpha)}_N=(1-\alpha)^{-1}\ln\sum_n p^\alpha(n)$ the R\'enyi
number entropy, generalizing Eq.~\eqref{Snum}. Any additional
entanglement is due to different configurations in each particle
sector having finite probability and is thus part of what we call the
R\'enyi configurational entropy.

\paragraph{System.---$\!\!\!\!$}
To be concrete, let us consider a half-filled fermionic model 
\begin{equation}
\label{tV}
H = -J\sum_j (c^\dagger_j c_{j+1} +h.c.)+\sum_j D_j n_j +V\sum_j n_j n_{j+1} \, ,
\end{equation}
with nearest-neighbor hopping amplitude $J$, interaction $V$, and
onsite disorder $D_j\in [-D/2,D/2]$. Here $n_j=c^\dagger_j c_j$ is the
particle number at site $j$. Using a Jordan-Wigner transformation,
this model can be mapped onto a spin-$1/2$ XXZ chain with magnetic
field disorder.  For $V=2J$, in particular, one obtains the isotropic
Heisenberg model which is the most studied system to investigate MBL
physics. We set $J=1$ and $\hbar=1$ in the following.

\paragraph{Thermalization.---$\!\!\!\!$}
If such a system after a quantum quench thermalizes
to a high temperature state, then a region of size $2\ell$ will
contain $\ell$ particles on average and every arrangement of particles
will approximately have equal probability. If we now cut the
thermalized region in half, then the probability to find $n$ particles
in one half is $p(n)={{\ell}\choose{n}}
{{\ell}\choose{\ell-n}}/{{2\ell}\choose{\ell}}$. For large $n,\ell$
this distribution can be approximated by a continuous distribution and
one finds for all R\'enyi number entropies (including $\alpha\to 1$) in
the ergodic case $S^{(\alpha)}_N = \mbox{const}
+\frac{1}{2}\ln\ell$ with $S^{(\alpha)}_N > S^{(\alpha+1)}_N$
\cite{KieferUnanyan1}. If the excitations in the system spread as
$t^\nu$ after the quench then the thermalized regions have size
$\ell\sim t^\nu$ and we obtain
\begin{equation}
\label{ergodic}
S^{(\alpha)}_N(t) = \mbox{const} +\frac{\nu}{2}\ln t \, .
\end{equation}

\paragraph{Localization.---$\!\!\!\!$}
The presence of disorder (i.e. $D\ne 0$) can prevent thermalization
and lead to localized states.  The simple scaling argument why free
particles ($V=0$) on a lattice with short-range hoppings become
localized for strong potential disorder works as follows
\cite{Anderson58,AbrahamsAnderson,AndersonLocalization}: A real
hopping process requires a resonance, i.e., an energy matching between
the two sites involved in the hopping process. The smallest mismatch
in energy on a subsystem of volume $\ell^d$ decreases as $\ell^{-2d}$
in $d$ dimensions on average. The transport between quasi-degenerate
states needs on the order of $n\sim \ell$ hopping processes and the
amplitude for such a virtual $n$-site hopping process falls of
exponentially with distance. Therefore distant resonances have a
vanishingly small probability to proliferate and to delocalize the
system. A non-interacting system at sufficiently strong disorder will
therefore be in an AL phase and both $S$ and $S_N$ will saturate. In
one dimension, even arbitrarily weak disorder is sufficient to
localize all states. The crucial question then is what influence
interactions have on the probability of distant resonances.

If the model \eqref{tV} is in an AL phase for $V=0$, a localized basis
$\{|\psi_l\rangle\}$ exists such that the non-interacting Hamiltonian
becomes diagonal, $H_0=\sum_l\varepsilon_l \eta_l=\sum_l\varepsilon_l
d^\dagger_l d_l$. We can transform
\eqref{tV} to this localized basis using $c_j^\dagger
= \sum_l
\langle \psi_l|\phi_j\rangle d_l^\dagger$, where $|\phi_j\rangle$ is the 
original Wannier basis. Here $l$ can be understood as the index of the
site around which the localized single-particle wavefunction is
centered, i.e., $|\langle \psi_l|\phi_j\rangle|^2\sim
\exp(-|l-j|/\xi_{\textrm{loc}})$ where $\xi_{\textrm{loc}}$ is the localization 
length. If we transform the interaction part to the new basis, we find
contributions describing density-density interactions between
localized orbitals as well as hopping processes between these
orbitals. The density-density part is given by
$H_{\textrm{int}}^{(1)} = \sum_{l,l'} J_{ll'} \eta_l \eta_{l'}$
with an amplitude which decays exponentially with distance between the
orbitals, $J_{ll'}\sim V\exp(|l-l'|/\xi_{\textrm{loc}})$. If this
would be the only relevant correction due to interactions, then
particles would remain localized with $H_{\textrm{int}}^{(1)}$ causing
a logarithmic buildup of configurational entanglement. However, the
interaction also leads to a correlated hopping between the single-particle
orbitals $\vert \psi_l\rangle$
\begin{equation}
\label{int2}
H_{\textrm{int}}^{(2)} = \sum_{l,l',k,k'} K_{ll'kk'} d^\dagger_ld_{l'}d^\dagger_kd_{k'}
\end{equation}
with unequal lattice sites and exponentially decaying amplitude
$K_{ll'kk'}$. Similar to the AL case, one then has to consider the
possibility of resonances destroying localization. In contrast,
hopping processes are now long-ranged so that both direct and virtual
transitions to distant sites are possible. The smallest expected
average mismatch in energy, $\Delta\varepsilon =
\varepsilon_l-\varepsilon_{l'}+\varepsilon_k-\varepsilon_{k'}$, on a
subsystem of length $\ell$ now decreases as $\ell^{-4}$. Without taking
the renormalization of the bare energies $\varepsilon_l$ into account,
one would thus still conclude that distant resonances do not
proliferate. On the other hand, numerical and experimental data
\cite{SchreiberHodgman} indicate that for small disorder 
interactions do destroy the localized phase. I.e., in this case energy
renormalizations do seem to lead to a proliferation of resonant
hopping processes. For strong disorder, on the other hand, it has been
argued that the processes \eqref{int2} are irrelevant and the
particles are localized
\cite{RosMuellerScardicchio,Imbrie2016}. However, these results are
based on approximations. The proof of MBL for weak interactions in
Ref.~\onlinecite{Imbrie2016}, in particular, is based on an assumption
about limited level attraction in the statistics of energy
eigenvalues. 

\paragraph{Numerical results.---$\!\!\!\!$}
Since the question about the relevance of resonances ultimately cannot be decided analytically, we investigate the number
entropy for the model \eqref{tV} by exact diagonalization (ED). We
concentrate on $V/J = 2$ corresponding to the isotropic Heisenberg
model. In our notation, the critical coupling for the transition from
the ergodic into the MBL phase is $D_c/J\approx 14$
\cite{PalHuse,Luitz1}. We study quenches starting from half-filled 
random product states. If not stated otherwise the data shown for
$L\leq 18$ are obtained by standard full diagonalizations of the
Hamiltonian, averaged over 10000 disorder realizations for $L\leq 14$
and 3000 realizations for $L>14$, while a second order Trotter-Suzuki
decomposition of the time evolution operator is used for $L=24$, see
App.~A for details.

\begin{figure}
	\includegraphics*[width=1\columnwidth]{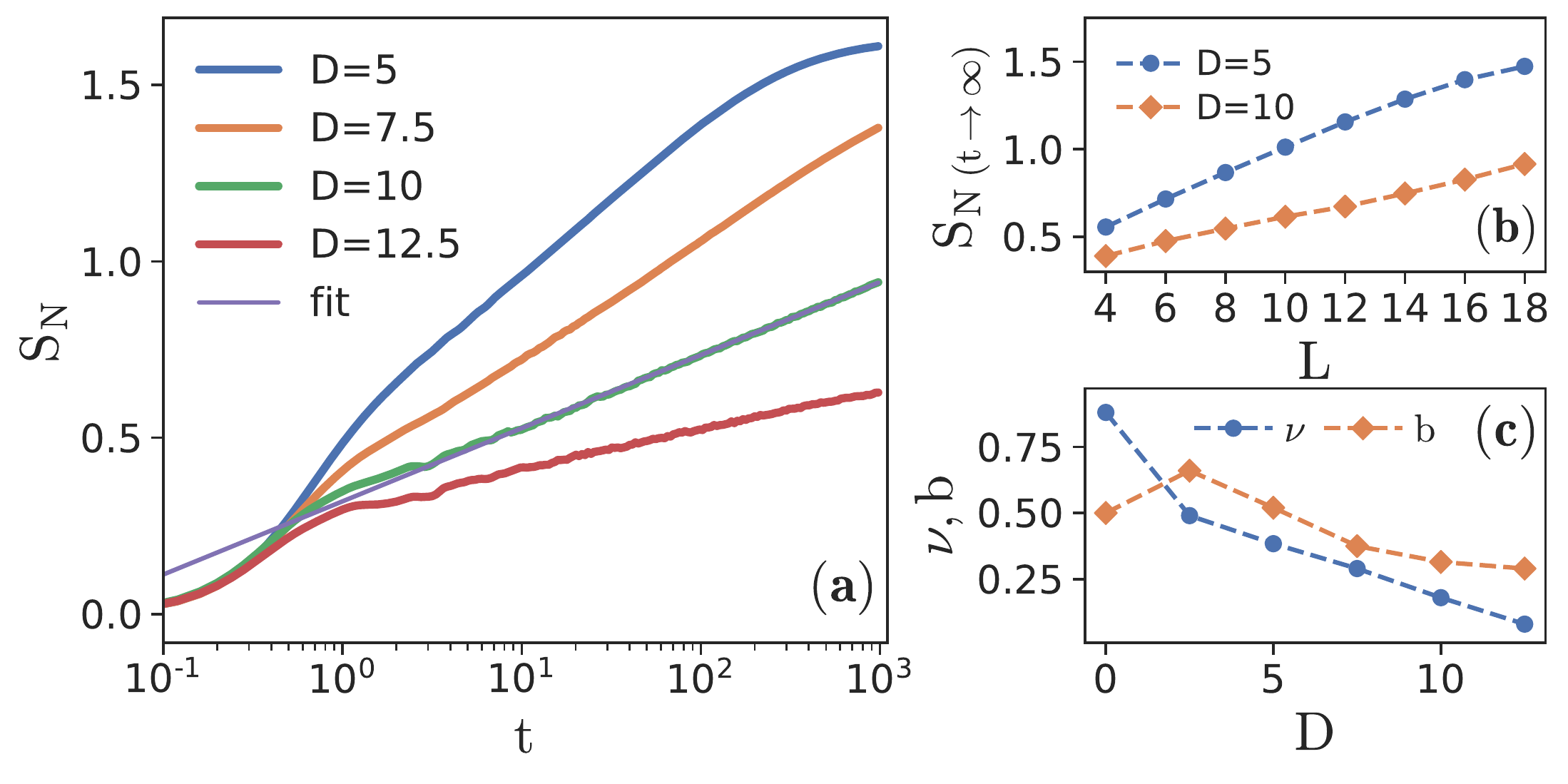}
	\caption{$S_N$ for $D<D_c\approx 14$: (a) $S_N(t)$ for $L=24$,
	with 500 disorder realizations and a logarithmic fit, $S_N
	=\frac{\nu}{2} \ln t +b$, for $D=10$ with $\nu=0.18,
	b=0.32$. (b) $S_N(t\to\infty)$ for different system sizes $L$. (c) Prefactors $\nu$ and constants $b$ of the
	logarithmic fits as a function of $D$ for $L=24$.}
\label{Fig1}
\end{figure}

Let us first consider the regime $D<D_c$ where there is consensus that
the system is ergodic. ED \cite{AgarwalGopalakrishnan,BarLevCohen},
large-scale density-matrix renormalization group (DMRG) calculations
\cite{ZnidaricScardicchio,EnssAndraschkoSirker}, and phenomenological
numerical renormalization groups \cite{PotterVasseurPRX,VoskHusePRX}
furthermore find subdiffusive transport either all the way down to
zero disorder or up to a second critical disorder below which
transport becomes diffusive. In contrast to the linear-in-time
spreading of excitations in the clean case, it now takes time $t\sim
\ell^{1/\nu}$ for excitations to spread across a region of length
$\ell$ with $\nu=1/2$ corresponding to diffusion. We therefore expect
$S\sim \ell\sim t^\nu$ with $S_N$ given by Eq.~\eqref{ergodic}. This
scaling of $S(t)$ in the ergodic regime is consistent with DMRG
calculations for infinite chains with binary disorder
\cite{EnssAndraschkoSirker} and with ED
\cite{BarLevCohen} for box disorder. In Fig.~\ref{Fig1}, results for
the number entropy of model \eqref{tV} at various disorder strengths
$D<D_c$ are shown. Here we consider systems of length $L$ with open
boundary conditions which are split into two equal halfs, $\ell=L/2$.
%
%
\begin{figure}
	\includegraphics*[width=1\columnwidth]{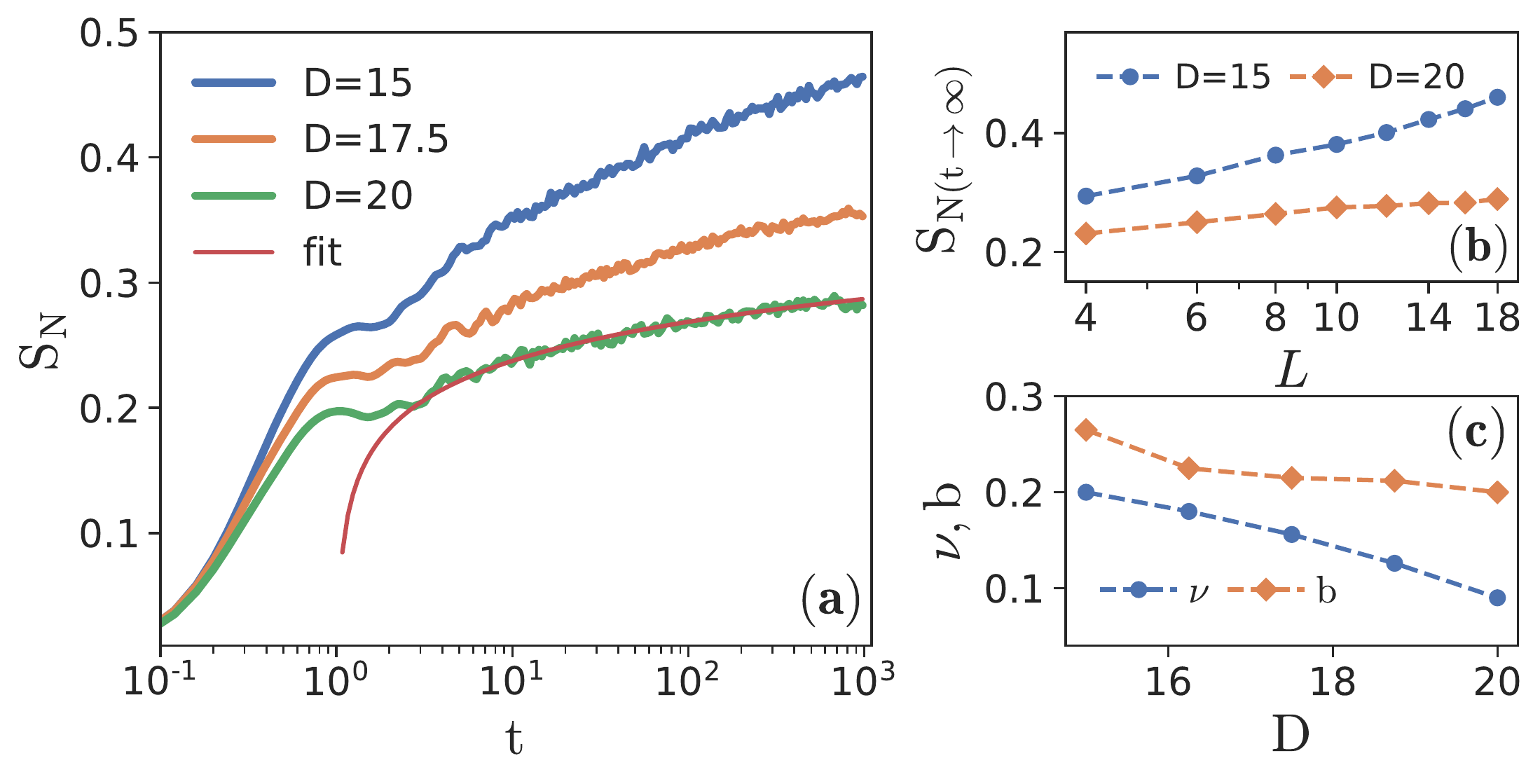}
	\caption{$S_N$ for $D>D_c$: (a) $S_N(t)$ for $L=24$ and double
		logarithmic fit, $S_N =\frac{\nu}{2} \ln\ln t +b$, for $D=20$
		with $\nu=0.09, b=0.20$. (b) $S_N(t\to\infty)$ for different
		system sizes $L$. (c) Prefactor $\nu$ and constant $b$
		of the double logarithmic fits as a function of $D$ for
		$L=24$, see also App.~B.}
	\label{Fig2}
\end{figure}
%
We find that $S_N(t)$ grows logarithmically consistent with
Eq.~\eqref{ergodic} and thus ergodic behavior. This is also supported
by the close to linear scaling of the saturation value $S_N(t\to\infty)$
with system size. Finally, we note that the prefactor $\nu$ decreases
continuously as a function of disorder $D$ and appears to approach
zero for $D\to D_c$. The results for the number entropy are
qualitatively consistent with previous results for the scaling of the
current \cite{ZnidaricScardicchio} and of the bipartite particle
number fluctuations $\Delta n$
\cite{BardarsonPollmann,SinghBardarsonPollmann}. 

Turning to the case $D>D_c$, it is expected that it then takes time
$t\sim\e^\ell$ to entangle regions over a distance $\ell$. The
resulting scaling of the von-Neumann entropy $S\sim\ln t$ has been
demonstrated already by various methods and for a number of different
models and our results are consistent with such a scaling as well. Our
main new result are the data for the number entropy presented in
Fig.~\ref{Fig2}.

We find that the number entropy continues to increase as
$S_N\sim\ln\ln t$ and that the saturation value continues to grow as a
function of length as in the ergodic case $D<D_c$, however, now only
approximately logarithmically. For the numerically accessible times
and lengths we find no indications for a saturation of the number
entropy as would be expected if the system is localized. Note that
$S_N\sim\ln\ln t$ is exactly the scaling which is anticipated if the
system is ultimately ergodic and $t\sim\e^\ell$ is not only the
relevant scaling for the buildup of configurational entanglement but
also for the spreading of particles (see derivation of
Eq.~\eqref{ergodic}). As a function of disorder strength $D$ we find
that the prefactor $\nu$ of the double logarithmic growth is
decreasing continuously. There are no indications for a sharp
transition. Let us also comment on the bipartite particle fluctuations
$\Delta n$ investigated previously
\cite{BardarsonPollmann,SinghBardarsonPollmann}. Our results (not shown) 
are consistent with $\Delta n(t)$ growing without bounds and $\Delta
n(t\to\infty,L)$ increasing with increasing system size $L$.

\begin{figure}
	\includegraphics*[width=1\columnwidth]{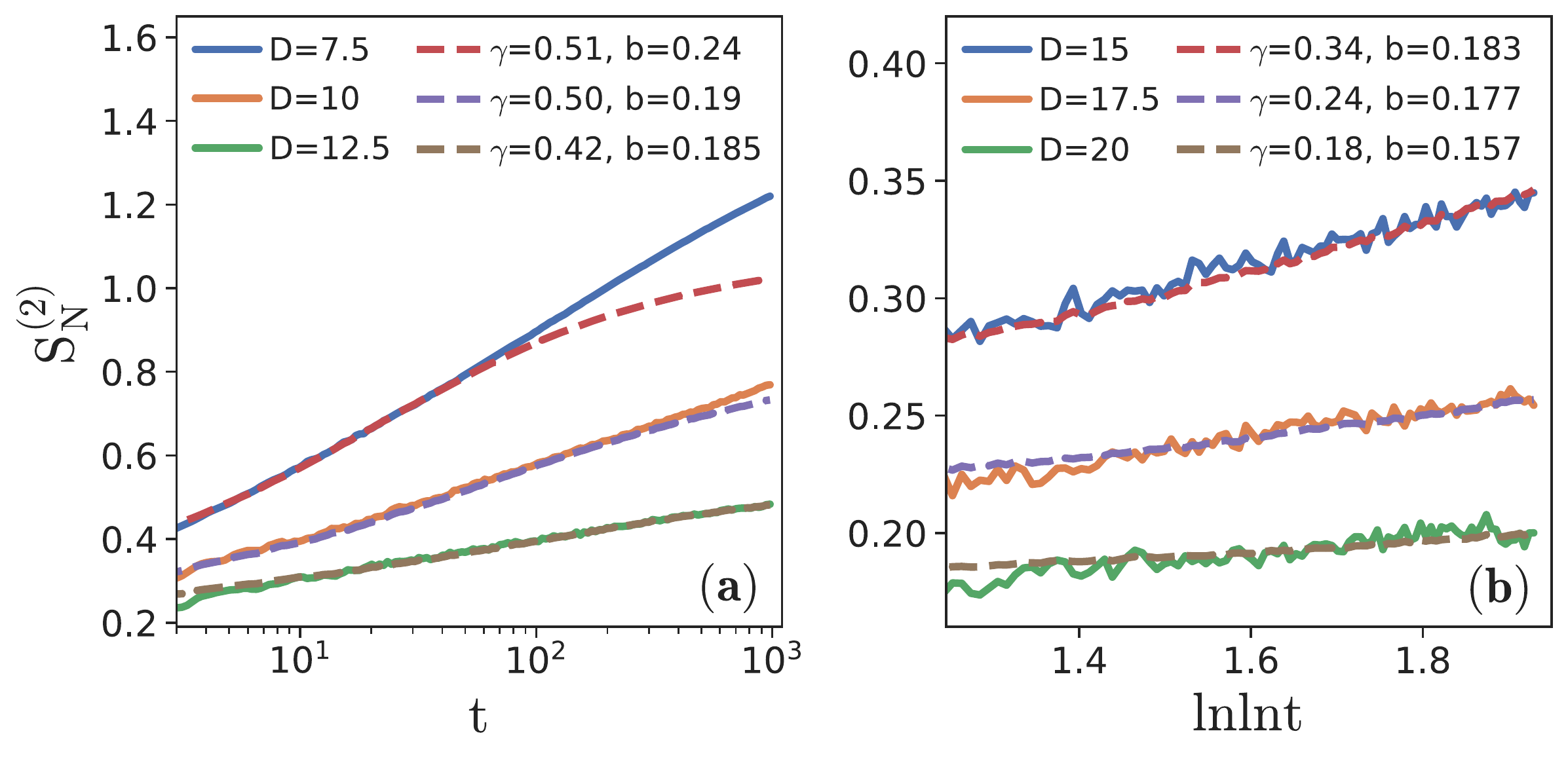}
	\caption{$S_N^{(2)}$ and bound \eqref{bound} for (a) $D<D_c$,
	and (b) $D>D_c$. The renormalization parameter $\gamma$
	appears to decrease monotonically with increasing $D$.}
\label{Fig3}
\end{figure}

To provide further support for an unbounded growth of the number
entropy, we now show that the numerical results are consistent with a
relation recently proven in the non-interacting case
\cite{KieferUnanyan1}. There we found that
\begin{equation}
\label{bound}
S_N^{(2)} \geq \gamma\left\{\frac{S^{(2)}}{2\ln 2} -\ln\left[I_0\left(\frac{S^{(2)}}{2\ln 2}\right)\right]\right\} +b
\end{equation}
provides a tight bound with $\gamma=1$, $b=0$, and $I_0$ being the
modified Bessel function. I.e., a growth of the second R\'enyi entropy
$S^{(2)}$ is always accompanied by a growth, albeit logarithmically
slower, of the corresponding number entropy $S_N^{(2)}$. In
Fig.~\ref{Fig3} we show that this bound with a renormalized $\gamma$
(and curves shifted by $b>0$ for ease of presentation) appears to
remain valid in the interacting case {\it both} for $D<D_c$ and
$D>D_c$.
%
%
Note that in the interacting case, i.e. $V \ne 0$, $\gamma$ appears to
decrease continuously with increasing $D$ but does not show
indications of a sharp transition.

Finally, we want to consider a system with very strong disorder to
check whether the increase of the number entropy is transient. To this
end, we consider the model \eqref{tV} with binary disorder
$D_j\in\{-D/2,D/2\}$. For $D\to\infty$ this will result in finite
segments which are coupled by the interaction term but not by hopping
processes. We, furthermore, limit the size of segments with equal
potential $\pm D/2$ to four lattice sites. In this case, the disorder
is no longer uncorrelated but this should only help in reducing the
time scale where $S_N^{(\alpha)}$ potentially saturates.
%
\begin{figure}
	\includegraphics*[width=1\columnwidth]{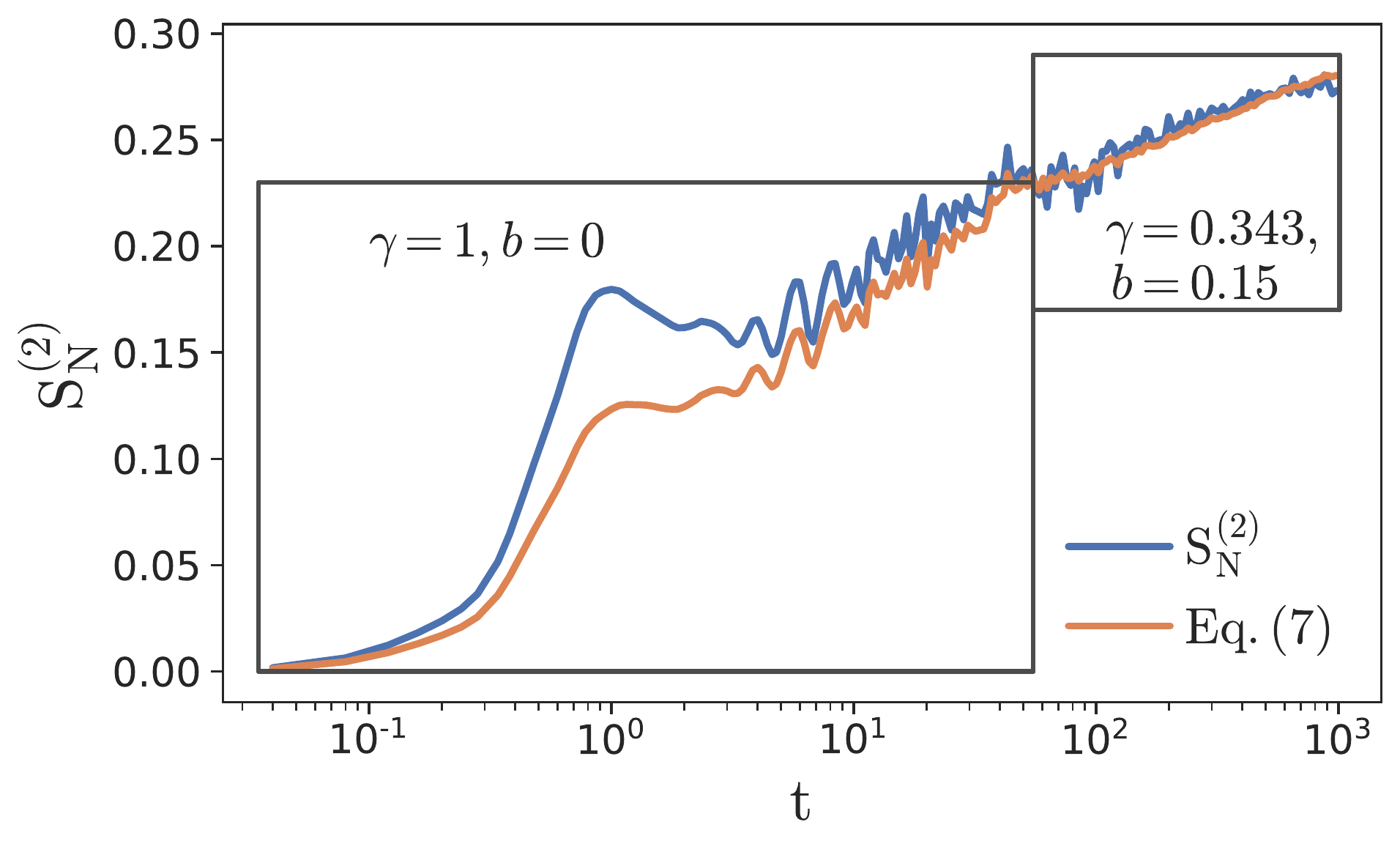}
	\caption{$S_N^{(2)}$ for model \eqref{tV} with binary disorder
	$D=20$ and segments with equal $D$ limited to four sites
	(1000 disorder realizations). For $t<t_\textrm{Th}$ an
	unrenormalized bound ($\gamma=1$) holds while $\gamma$ is
	renormalized for $t>t_\textrm{Th}$, see text.}
\label{Fig5}
\end{figure}
%
Numerically, we find that the bound \eqref{bound} also holds in this
case, see Fig.~\ref{Fig5}. For $t\lesssim t_\textrm{Th}$, where $t_\textrm{Th}\sim
\exp(D/\Omega) L^2$ \cite{SuntajsBonca} 
is the Thouless time with $\Omega$ being a constant, the original
bound in the non-interacting case ($\gamma=1$) holds, while a
renormalized bound holds for longer times (see also App.~C). In
the thermodynamic limit, $S_N^{(2)}$ thus appears to grow without
bounds in this model as well.

\paragraph{Conclusions.---$\!\!\!\!$}
The slow increase of the number entropy $S_N\sim\ln\ln t$ and the
increase of the saturation value as a function of system size, found in our numerical simulations, are not
expected in an MBL phase. There are at least two different possible
interpretations of these data. First, it cannot be excluded that the
observed behavior after all is transient and that $S_N$ in the
thermodynamic limit does saturate at very long times. While this
interpretation would not challenge the established phenomenology of
MBL phases, it is then an open question to understand the origin of
such a long-time transient behavior as well as the time and length
scales where particle fluctuations ultimately cease to grow. While
finite-size effects have been suggested to strongly affect numerical
studies of MBL
\cite{Abaninrecent}, we note that in contrast to
Ref.~\onlinecite{SuntajsBonca} our data---which also challenge the
established MBL phenomenology---are obtained at strong
disorder. Second, it is possible that hopping processes introduced by
the interaction term \eqref{int2} are relevant and resonances do
exist. A possible scenario would be that for $D>D_c$ the dynamic
scaling $t\sim\e^\ell$ does hold, leading to a logarithmic growth of
the entanglement entropy but that the same dynamical scaling also
holds for the spreading of particles resulting in an unbounded growth
$S_N\sim\ln\ln t$. While this implies that the system is ultimately
ergodic at very long time scales, it will not drastically alter the
behavior on experimentally accessible time scales: MBL systems would
still be good quantum memories and the Hamiltonian
\eqref{Heff} an effective description.

\acknowledgments
J.S. acknowledges support by the Natural Sciences and Engineering
Research Council (NSERC, Canada) and by the Deutsche
Forschungsgemeinschaft (DFG) via Research Unit FOR 2316. We are
grateful for the computing resources and support provided by Compute
Canada and Westgrid. M.K., R.U. and M.F. acknowledge financial support
from the Deutsche Forschungsgemeinschaft (DFG) via SFB TR 185, project
number 277625399. The simulations were (partly) executed on the high
performance cluster "Elwetritsch" at the University of Kaiserslautern
which is part of the "Alliance of High Performance Computing
Rheinland-Pfalz" (AHRP). We kindly acknowledge the support of RHRK.
M. K. would like to thank J. L\'eonard and M. Greiner for hospitality
and fruitful discussions and J. Otterbach for advice for the
optimization of algorithms to GPU's.

\appendix

\section{Appendix A: Numerical methods}
\label{AppA}
\begin{figure}[h]
\includegraphics[width=0.9\columnwidth]{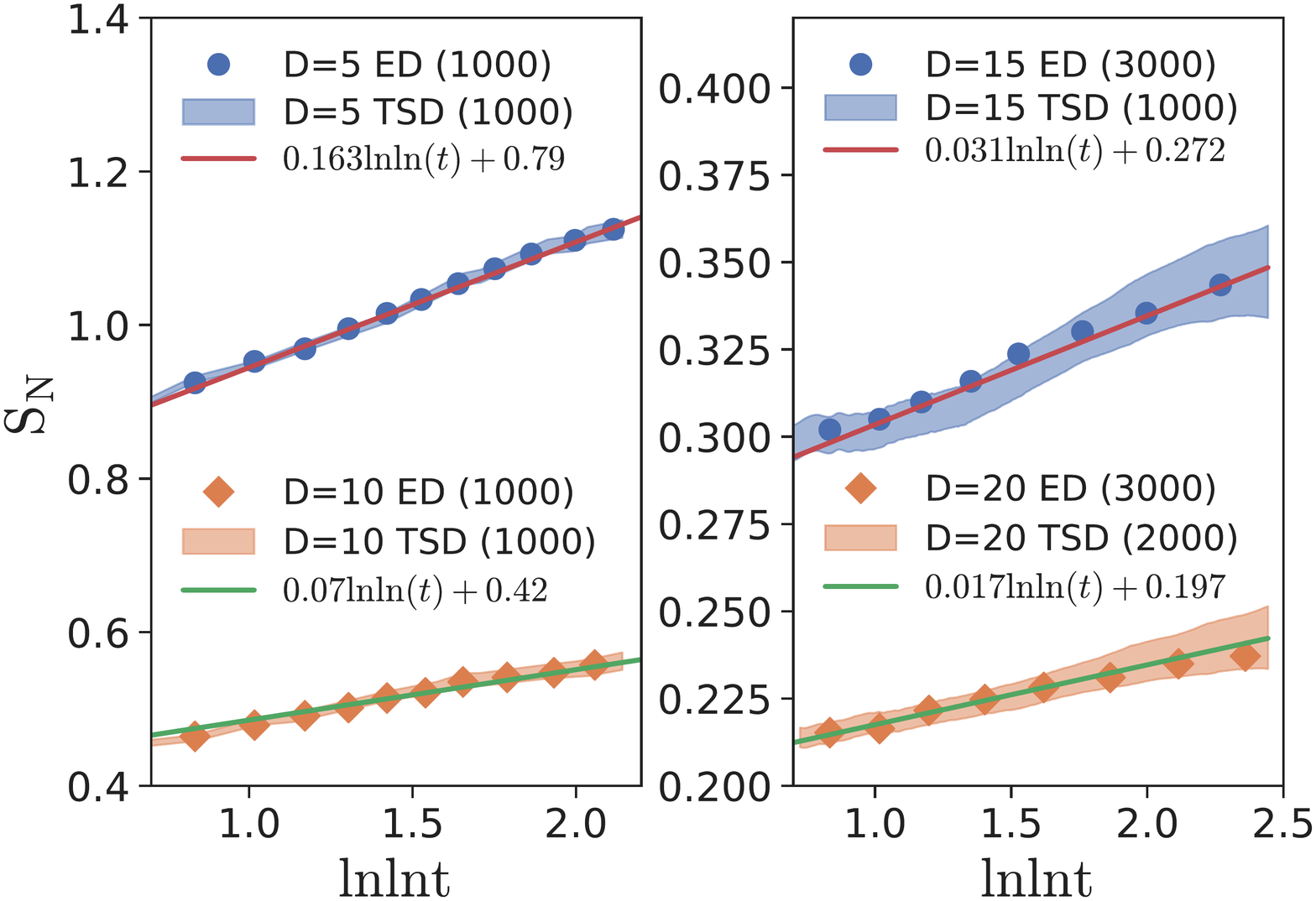} 
\caption{$S_N$ for $D >D_c\approx 2$ with $V=0.2$ and $L=16$: 
(a) $S_N(t)$ computed by ED (dots) compared to second order TSD
(shaded areas), averaged over 1000 disorder realizations. The lines
show fits $\sim \ln\ln t$. (b) The same as in (a) but for stronger
disorder and different numbers of realizations, see legend. The shaded
areas represent a rolling average which contain 99\% of the TSD data
with errorbars.}
\label{compare_trotter_ed}
\end{figure}

We first discuss the numerical methods used in more detail. If not
stated otherwise, our results are obtained by standard ED methods for
open boundary conditions for system sizes $L\leq 18$ using particle
number conservation. For system sizes $L> 18$, we employ a second
order Trotter-Suzuki decomposition (TSD)\cite{Trotter,Suzuki1,
Suzuki2} of the time evolution operator to evolve the full many-body
state
\begin{equation}
\mathrm{e}^{-\mathrm{i}(\hat{A}+\hat{B})\delta t} = \mathrm{e}^{-\mathrm{i}\frac{\hat{A}\delta t}{2}}\mathrm{e}^{-\mathrm{i}\hat{B}\delta t}\mathrm{e}^{-\mathrm{i}\frac{\hat{A}\delta t}{2}} + \mathcal{O}(\delta t^3),
\end{equation}
where $\hat{A}, \hat{B}$ are non-commuting operators given by the even
and odd parts of the Hamiltonian. Since we apply the decomposed
time-evolution operator to the full state, no truncations are needed
as, for example, in TEBD algorithms \cite{SerbynPapicPRX}. The cumulative error in the time evolution of a system using a second order TSD is thus quadratic in $\delta t$. Therefore, the simulation time is restricted by the step size of the TSD. We use $\delta
t=0.01$ for computations with $D<D_c$ and $\delta t=0.005$ for
computations with $D>D_c$, because the latter is the regime we are
most interested in. In order to reach times beyond $10^4$ hopping
amplitudes, see Fig.~\ref{compare_trotter_ed}, higher-order
Trotter-Suzuki decompositions would be more efficient. The second
order TSD is, however, sufficient for our purposes. The TSD method can
be easily parallelized on a large array of graphical computing units
(GPU's), making system sizes of up to $L=24$ accessible.

\section{Appendix B: Scaling deep in the MBL phase}
\label{AppB}
\begin{figure}[h]
\includegraphics[width=0.9\columnwidth]{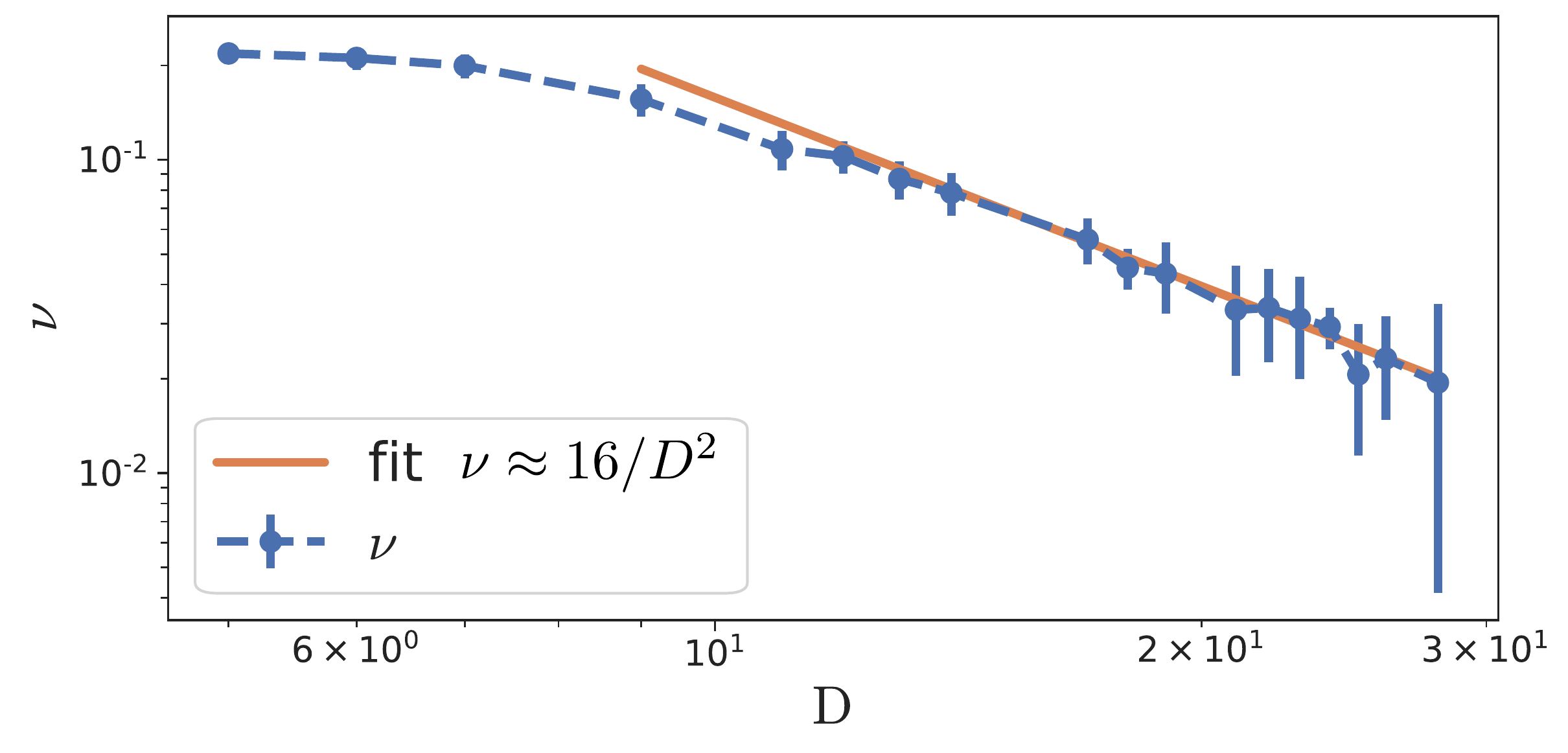} 
\caption{Parameter $\nu$ extracted from fits $S_N\approx\frac{\nu}{2}\ln\ln t+b$ as function of disorder strength $D$ for $V=0.2$ and $L=12$. Error bars
represent uncertainties related to varying $\nu$, $b$, and the fit
interval.}
\label{fig2}
\end{figure}

For weak interactions it has been shown that the critical disorder
strength $D_c$, needed to localize a many-body system, is small. In the
following, we consider $V=0.2$ where the critical disorder strength
has been estimated to be $D_c\approx 2$, see
Ref.~\cite{BarLevCohen}. We note, however, that it has been argued
that for very large system sizes $L>100$ the critical disorder
strength might actually be about a factor of $2$ larger
\cite{Doggen2018}. We therefore investigate the behavior of the
prefactor $\nu$ when fitting the number entropy according to $S_N
\sim \frac{\nu}{2}
\ln\ln t$ for disorder strengths $D\in [5,30]$. The larger MBL phase 
expected at weak interactions gives us better access to the scaling
behavior of $S_N$. As can be seen from Fig.~\ref{fig2}, $\nu$ is
decreasing with increasing disorder but not faster than a power law,
even for very strong disorder about 15 times the critical disorder
strength. Most importantly, we do not find any indications for a sudden
drop or a phase transition.
%

\section{Appendix C: Renormalization of bounds and Thouless time}
\label{AppC}
\begin{figure}[h]
\includegraphics[width=0.49\columnwidth]{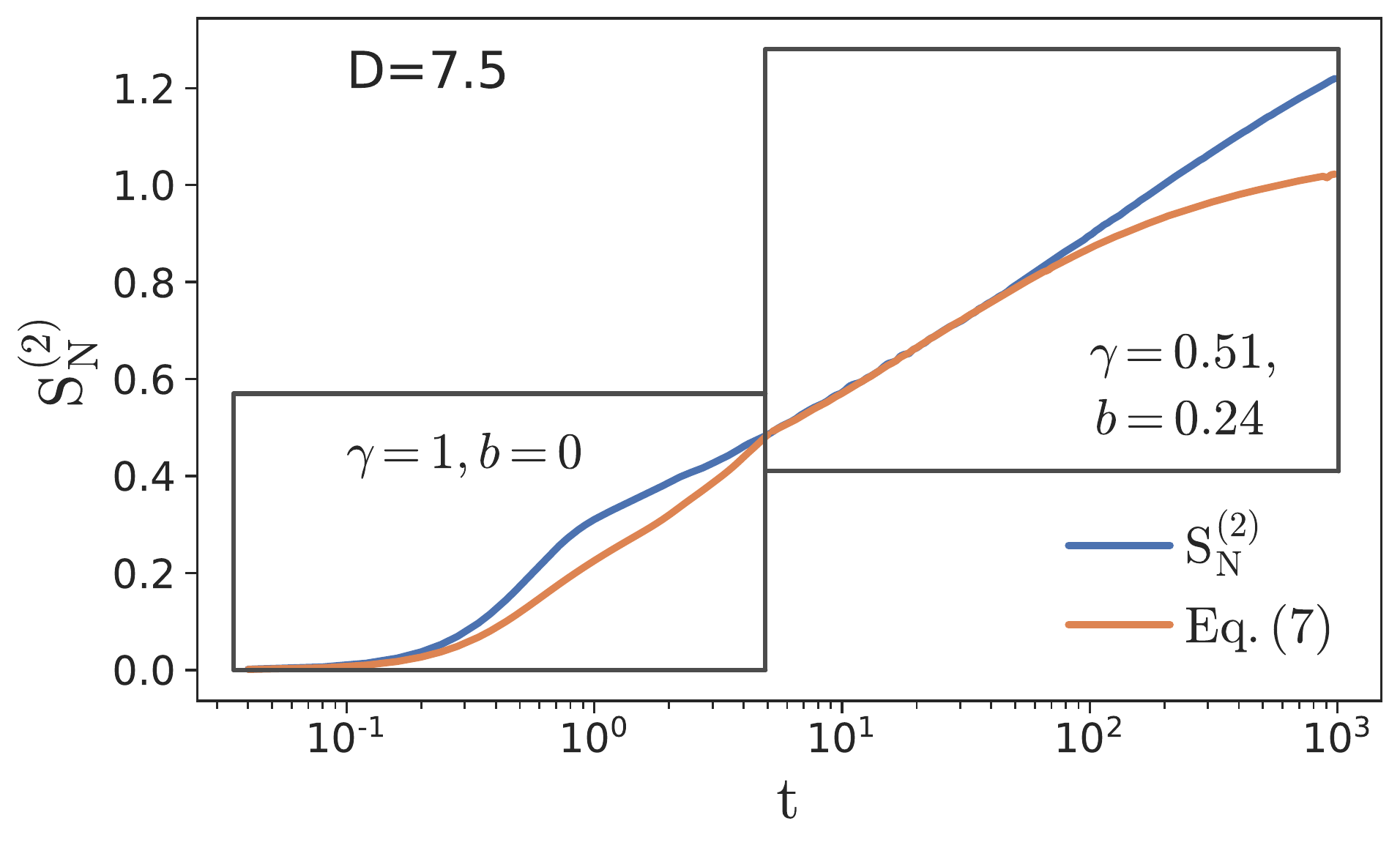} 
\includegraphics[width=0.49\columnwidth]{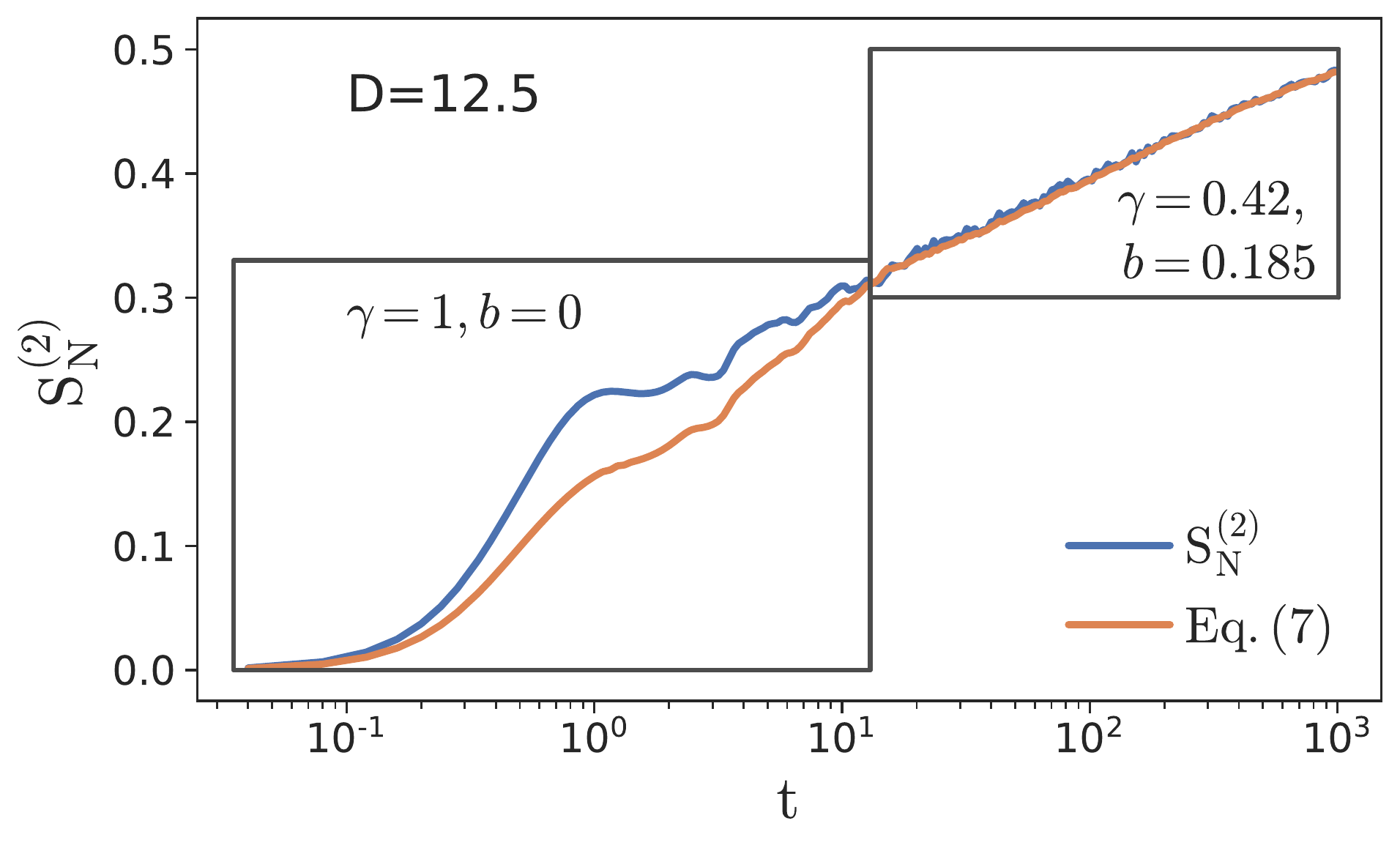} 
\includegraphics[width=0.49\columnwidth]{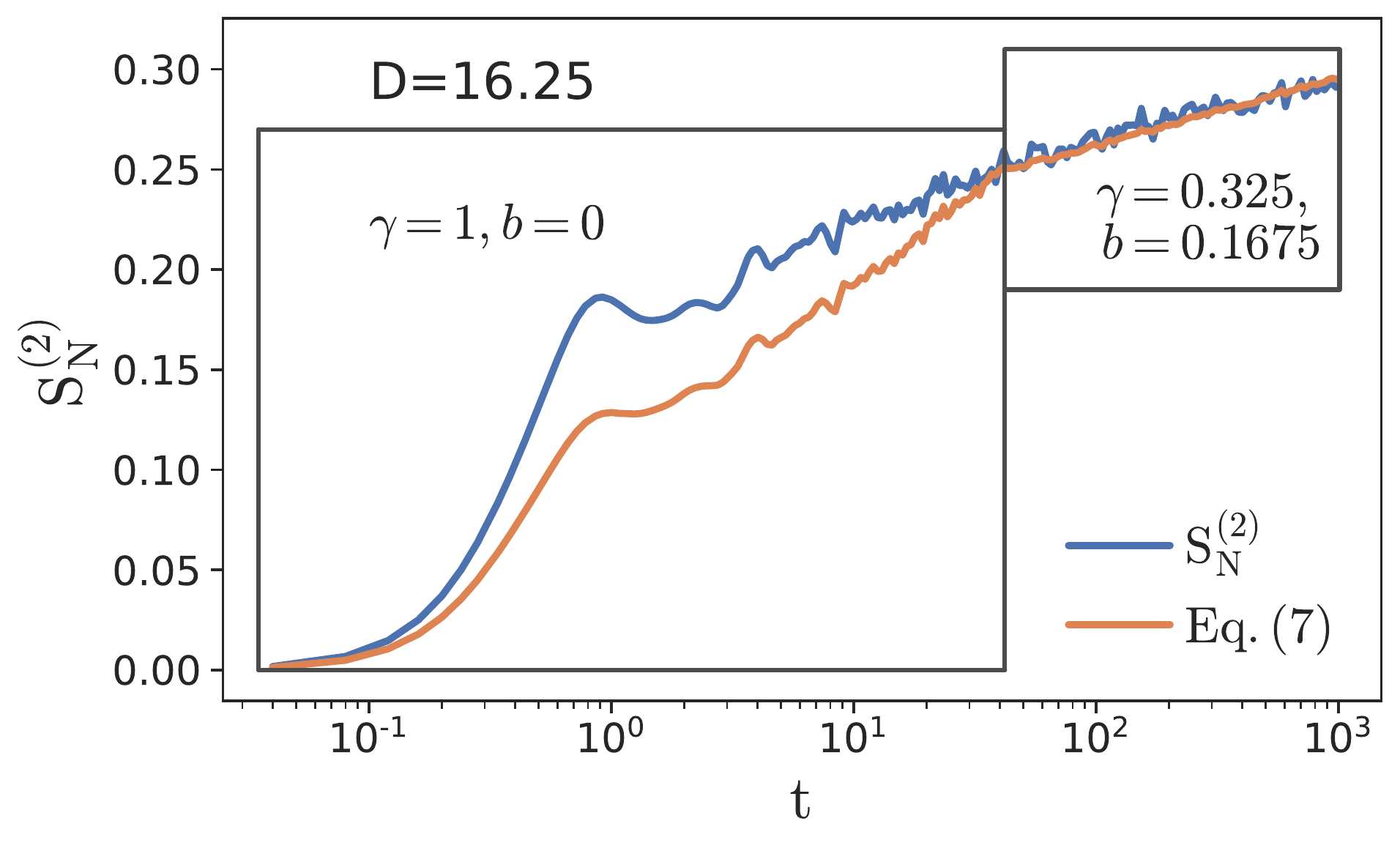} 
\includegraphics[width=0.49\columnwidth]{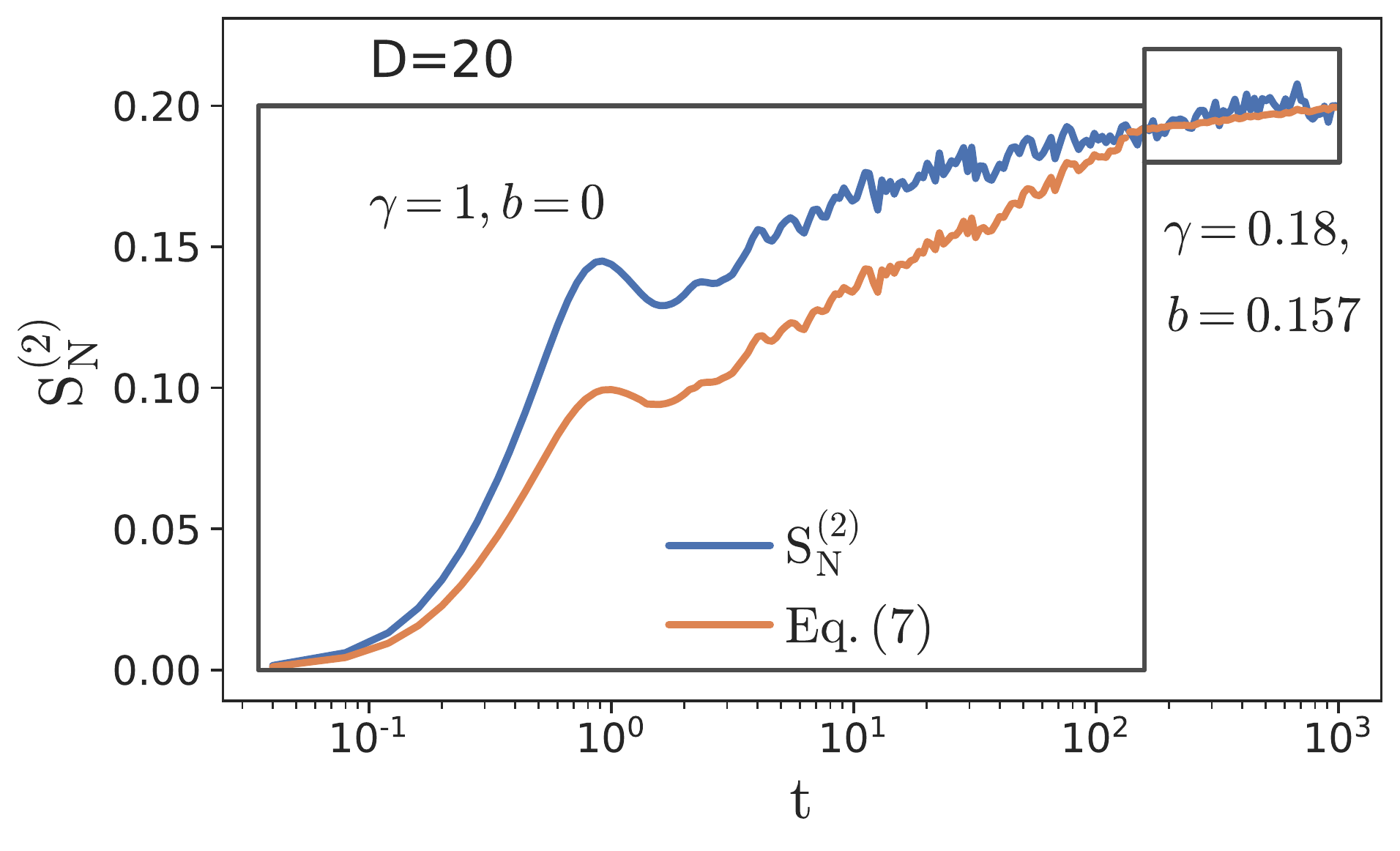} 
\caption{Lower bound, Eq.(\ref{bound}), and number entropy $S^{(2)}_N$ for $V=2$. 
The boxes indicate where the bound with $\gamma=1$ and $b=0$ crosses
$S^{(2)}_N$ for the first time.}
\label{compare_Dis}
\end{figure} 
In the main text, we have seen that the a lower bound for the number
entropy recently obtained for free fermions \cite{KieferUnanyan1}
appears to hold also in the interacting case but with parameters
renormalized by interactions and disorder. This bound is given by Eq.~\eqref{bound}.
%
%
For binary disorder, see Fig.~4 in the main text, we found that at
small times the free-fermion bound applies directly, while beyond a
certain time $t_\textrm{x}$ a renormalized value of $\gamma$ had to be
used. We now give further evidence for this behavior in the case of
box disorder and $V=2$, see Fig.~\ref{compare_Dis}, both for
$D<D_c\approx 14$ and $D>D_c$.  The figure shows that we can find
perfect fits in this case as well. We define the crossover time
$t_\textrm{x}$ as the time where Eq.~(\ref{bound}) with
unrenormalized parameters crosses $S_N^{(2)}$. For $t>t_\textrm{x}$,
we have to renormalize $\gamma$ in order for the bound to hold.

\begin{figure}[h]
	\includegraphics[width=0.9\columnwidth]{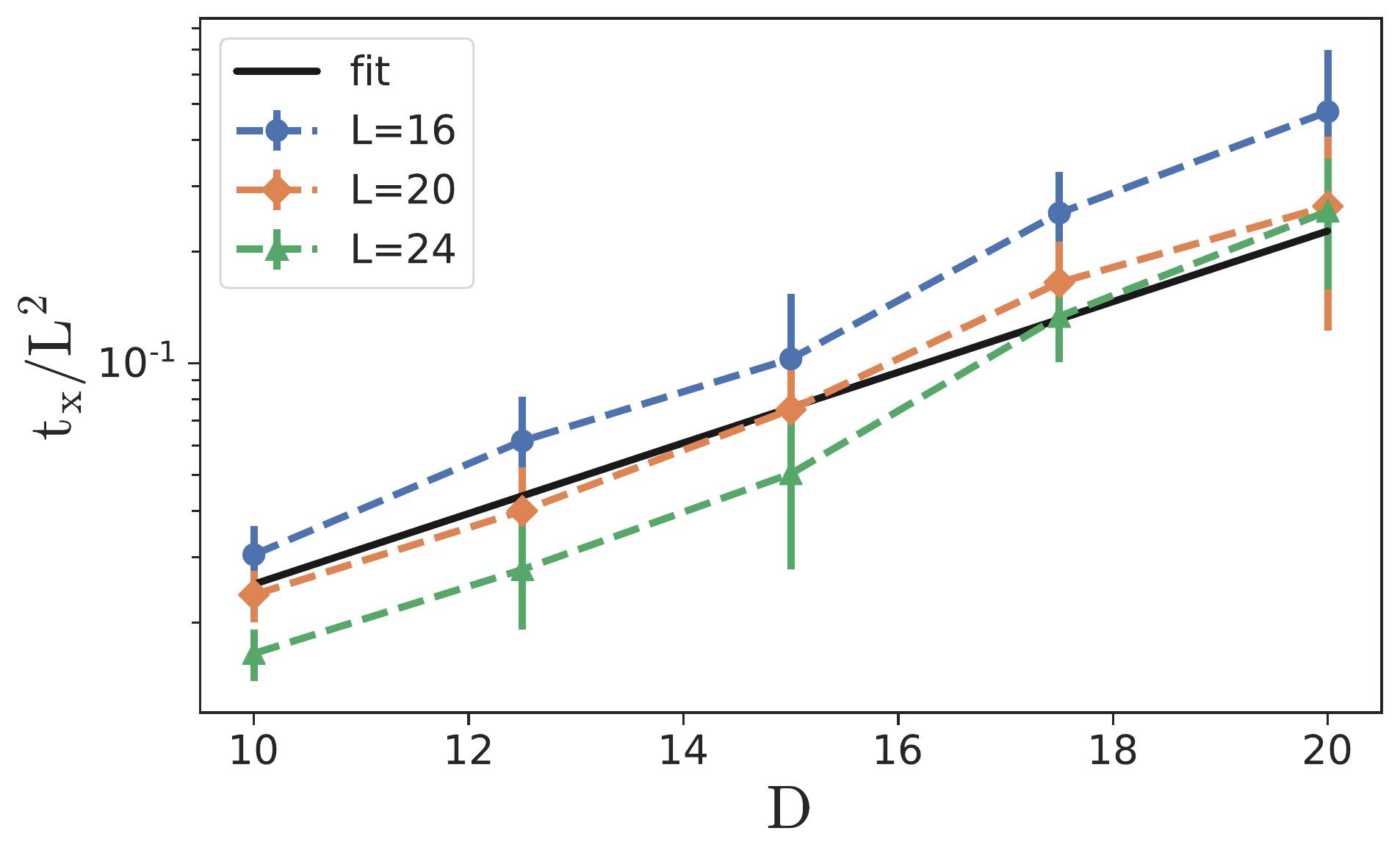}
	\caption{Scaling behavior of the crossover time $t_\textrm{x}$ between regimes with $\gamma=1$ and $\gamma < 1$. The fit shows the scaling behavior of the Thouless times \cite{SuntajsBonca}, $t_\textrm{Th}/L^2= t_0\mathrm{e}^{D/\Omega}$, where $t_0=2.84\times 10^{-3}$ is the characteristic time as in \cite{SuntajsBonca} and $\Omega=4.56$ is obtained from fitting the data. }
	\label{thouless}
\end{figure}
\noindent
We now argue that $t_\textrm{x}$ is related to the Thouless time
$t_\textrm{Th}$. To this end, we estimate the crossover time from the
data and analyze its scaling with systems size and disorder strength.
Note that in most cases we do not have a sharp crossing point in time,
so we give a range of values for $t_\textrm{x}$ using as a criteria that
the absolute value of the difference between the unrenormalized bound
and $S^{(2)}_N$ is less than $0.01$. Fig.~\ref{thouless} then
indicates that within these error margins, the crossover time
$t_\textrm{x}$ appears to show the scaling expected for the Thouless
time, $t_\textrm{Th}\sim L^2\mathrm{e}^{D}$ \cite{SuntajsBonca}.


\end{document}